\documentclass{article}

\usepackage[top=1.355in,
			bottom=1.355in,
			left=1.5in,
			right=1.5in]{geometry}
\usepackage{amsfonts,
            amssymb,
			amsmath,
			graphicx,
            subfig,
            booktabs,
			url}

\title{Data Assimilation: Addressing Spurious Correlations and Scalability Issues}

\author{Eric Crislip \thanks{Department of Statistics, The Ohio State University, escrisl@sandia.gov}, \and Moe Khalil \thanks{Sandia National Laboratories, mkhalil@sandia.gov}, \and Kyle Neal \thanks{Sandia National Laboratories, kneal@sandia.gov}}

\begin{document}

\maketitle

\begin{abstract}

Data assimilation in the form of the ensemble Kalman filter (EnKF) is commonly used to combine large-scale physics-based models and real-world observations of a quantity of interest. However, the EnKF is known to be adversely affected by spurious correlations when applied to Numerical Weather Prediction (NWP) models with high-dimensional state spaces. To mitigate the effect of spurious correlations, various localization methods have been proposed in the NWP literature. Unfortunately, the efficacy of and need for these methods in the setting of general PDE-based forward models with large spatial mesh sizes is not well-understood. We conducted a numerical data assimilation experiment on an example PDE model to demonstrate the presence of spurious correlations and need for localization methods outside of the context of NWP. In addition, we compared the computational efficiency of various EnKF implementations in high-dimensional settings for which theoretical complexity bounds are available but empirical costs are unclear.

\end{abstract}

\section{Motivation}    
Data assimilation is the process in which scientific or engineering models and observations are combined to create estimates of an unknown state that is evolving in time. The ensemble Kalman filter (EnKF) has emerged as the premiere choice of data assimilation technique in the context of expensive and large-scale dynamical systems. One such application area is Numerical Weather Prediction (NWP), where the EnKF has been studied extensively. However, the extension of methods developed in the context of NWP to broader PDE-based systems has been hitherto lacking. The objective of this paper is to investigate the application of the EnKF to general large-scale dynamical systems in two regards: we will examine both the phenomenon of spurious correlations as noted in the NWP literature and the performance of popular algorithms developed to reduce them, as well as compare the empirical computational costs of various EnKF analysis step formulations.

\section{Data assimilation}

    \subsection{Bayesian inference on dynamical systems}

    Let $\mathbf{u}_t \in \mathbb{R}^p$ denote the unknown state at time $t$. We model the time-evolution of $\mathbf{u}_t$ by
    \begin{equation}
        \mathbf{u}_t = \mathcal{M}_t(\mathbf{u}_{t-1}),
    \end{equation}
    where $\mathcal{M}_t (\cdot)$ is our \emph{forward model} which determines the \emph{evolution distribution} $p(\mathbf{u}_t | \mathbf{u}_{t-1})$. Implicitly, we make a Markovian assumption, where the current state depends only on the state at the previous time-step when conditioning on all previous states. If the initial state $\mathbf{u}_0$ is unknown, we may also define a prior distribution $p(\mathbf{u}_0)$ for it. We model measurements $\mathbf{y}_t \in \mathbb{R}^{q_t}$ of the unknown state by the expression
    \begin{equation}
        \mathbf{y}_t = \mathcal{H}_t(\mathbf{u}_t) + \boldsymbol \epsilon_t,
    \end{equation}
    where $\mathcal{H}_t(\cdot)$ is the observation operator with measurement noise $\boldsymbol \epsilon_t$, which is typically assumed to follow a Gaussian distribution. We will also assume measurements depend only on the current state and are independent of past and future measurements when conditioning on the true state.

    The goal of data assimilation is often to obtain the \emph{forecast distribution} $p(\mathbf{u}_t | \mathbf{y}_1, \ldots, \mathbf{y}_{t-1})$, which estimates the unknown state at a future time given current knowledge, and the \emph{analysis distribution} $p(\mathbf{u}_t | \mathbf{y}_1, \ldots, \mathbf{y}_t)$, which estimates the unknown state at the time a measurement is collected \cite{WIKLE20071}. 
    
    Under the stated conditional independence assumptions, we can acquire the forecast distribution by integration:
    \begin{equation}
        p(\mathbf{u}_t \mid \mathbf{y}_1, \ldots, \mathbf{y}_{t-1}) = \int p(\mathbf{u}_t \mid \mathbf{u}_{t-1}) p(\mathbf{u}_t \mid \mathbf{y}_1, \ldots, \mathbf{y}_{t-1}) d \mathbf{u}_{t-1}.
    \end{equation}
    The analysis distribution of $\mathbf{u}_t$ can then be computed by Bayes' rule:
    \begin{align}
        p(\mathbf{u_t} | \mathbf{y}_1, \ldots, \mathbf{y}_t) &= \frac{p(\mathbf{y}_t, \mathbf{u}_t \mid \mathbf{y}_1, \ldots, \mathbf{y}_{t-1})}{p(\mathbf{y_t})} \\
        &\propto p(\mathbf{y}_t, \mathbf{u}_t \mid \mathbf{y}_1, \ldots, \mathbf{y}_{t-1}) \\
        &= p(\mathbf{y}_t \mid \mathbf{u}_t) p(\mathbf{u}_t \mid \mathbf{y}_1, \ldots, \mathbf{y}_{t-1})
    \end{align}
    However, the desired distributions are rarely available in closed form and must be obtained through numerical means.

    \subsection{Kalman filter}

    Forecast and analysis distributions are available analytically when the forward model and observation model are both linear and all stochasticitiy is additive Gaussian, i.e.
    \begin{equation}
        \mathbf{u}_0 \sim \mathcal{N}(\boldsymbol \mu^f_0, \mathbf{P}^f_0),
    \end{equation}
    \begin{equation} \label{eq:linear observation operator}
        \mathbf{y}_t = \mathbf{H}_t \mathbf{u} + \boldsymbol \epsilon_t, \quad \mathbf{H}_t \in \mathbb{R}^{q_t \times p},
    \end{equation}
    \begin{equation}
        \boldsymbol \epsilon_t \sim \mathcal{N}(\mathbf{0}, \mathbf{R}_t),
    \end{equation}
    \begin{equation} \label{eq:ev operator nonlinear}
        \mathcal{M}_t(\mathbf{u}) = \mathbf{M}_t \mathbf{u} + \mathbf{q}_t, \quad \mathbf{M}_t \in \mathbb{R}^{p \times p},
    \end{equation}
    \begin{equation} \label{eq:gaussian model error}
        \mathbf{q}_t \sim \mathcal{N}(\mathbf{0},\boldsymbol \Lambda_t).
    \end{equation}
    In this setting, our Bayesian procedure reduces to the \emph{Kalman filter} \cite{kalman1960new}. When we have a Markovian assumption for the current state and assume measurements are independent when conditioning on the state variable, the analysis distribution is
    \begin{equation}
        \mathbf{u}_t \mid \mathbf{y}_1, \ldots, \mathbf{y}_{t-1} \sim \mathcal{N}(\boldsymbol \mu^f_t, \boldsymbol P^f_t),
    \end{equation} and the forecast distribution is
    \begin{equation}
        \mathbf{u}_t \mid \mathbf{y}_1, \ldots, \mathbf{y}_{t} \sim \mathcal{N}(\boldsymbol \mu^a_t, \boldsymbol P^a_t),
    \end{equation}
    where
    \begin{equation}
        \boldsymbol \mu^f_t = \mathbf{M}_t \boldsymbol \mu^a_{t-1},
    \end{equation}
    \begin{equation}
        \mathbf{P}^f_t = \mathbf{M}_t \mathbf{P}^a_{t-1} \mathbf{M}_t' + \boldsymbol \Lambda_t,
    \end{equation}
    \begin{equation}
        \mathbf{P}^a_t = (\mathbf{I} - \mathbf{K}_t \mathbf{H}_t) \mathbf{P}^f_{t},
    \end{equation}
    \begin{equation}
        \boldsymbol \mu^a_t = \boldsymbol \mu^f_t + \mathbf{K}_t (\mathbf{y}_t - \mathbf{H}_t \boldsymbol \mu_t^f).
    \end{equation}
    The matrix $\mathbf{K}_t = \mathbf{P}^f_t \mathbf{H}_t' (\mathbf{H}_t \mathbf{P}_t^f \mathbf{H}_t' + \mathbf{R}_t)^{-1}$ is known as the \emph{Kalman gain}.

    \subsection{Ensemble Kalman filter}
    Recall that the goal of data assimilation is to obtain the forecast and analysis distributions for the hidden state. When equations (\ref{eq:ev operator nonlinear}) and (\ref{eq:gaussian model error}) do not hold, the analytic expressions of the Kalman filter are invalid and the forecast and analysis distributions must be obtained by other means. A popular approach is to use ensemble-based methods, which approximate the forecast and analysis distributions through a discrete set of representative ensemble members (also known as particles). The \emph{ensemble Kalman filter} (EnKF) is one such method that accomodates nonlinear forward models through the use of particles, yet still utilizes analytic expressions for the analysis update \cite{evensen2006data}.

    The EnKF algorithm begins by generating $N$ ensemble members from the prior on the initial state, $\{\mathbf{u}_{0,i}\}_{i=1}^N \overset{iid}{\sim} p(\mathbf{u}_0)$. Evolving these ensemble members in time to the forecast distribution of $\mathbf{u}_1$ is straightforward. We simply pass the ensemble members through the forward model, $\mathbf{u}^f_{1,i} = \mathcal{M}_1(\mathbf{u}_{0,i})$ for $i=1,\ldots,N$. However, updating our forecast ensemble members for  $\mathbf{u}_1$ to the analysis distribution of $\mathbf{u}_1$ is more challenging. The first hurdle is the choice of algorithm. Multiple deterministic updating schemes exist for the EnKF, such as the Square Root Filter \cite{evensen2006data}. We will utilize a stochastic approach, which updates each ensemble member by
    \begin{equation} \label{eq:analysis updates}
        \mathbf{u}_{t,i}^a = \mathbf{u}_{t,i}^f + \mathbf{K}_t \big(\mathbf{y}_t + \boldsymbol \epsilon_{t,i} - \mathbf{H}_t\mathbf{u}^f_{t,i} \big), \quad i=1,\ldots,N, 
    \end{equation}
    \begin{equation}
        \{\mathbf{e}_{t,i}\}_{i=1}^N \overset{iid}{\sim} \mathcal{N}(\mathbf{0},\mathbf{R}_t)
    \end{equation}
    Once our ensemble members have been updated to the analysis distribution of $\mathbf{u}_1$, we can iterate between forecast evolutions and analysis updates for each subsequent timestep.

    Though the ability to accommodate nonlinear forward models is a strength of the EnKF, the algorithm still relies on a linear observation operator. If $\mathcal{H}_t$ is nonlinear, one option is to linearize it by taking $\mathbf{H}_t$ to be its Jacobian.

    In practice, we do not know the forecast the forecast mean $\boldsymbol \mu_t^f$ or covariance $\mathbf{P}^f_t$ and must approximate these moments by the ensemble estimates,
    \begin{equation}
        \hat{\boldsymbol \mu}_f^t = \frac{1}{N} \sum_{i=1}^N \mathbf{u}^f_{t,i},
    \end{equation}
    \begin{equation}
        \hat{\mathbf{P}}_f^t = \frac{1}{N-1} \sum_{i=1}^N (\mathbf{u}^f_{t,i} - \boldsymbol \mu_t^f)(\mathbf{u}^f_{t,i} - \boldsymbol \mu_t^f)',
    \end{equation}
    and use a plug-in estimator for the Kalman gain,
    \begin{equation}
        \hat{\mathbf{K}}_t = \hat{\mathbf{P}}^f_t \mathbf{H}_t' (\mathbf{H}_t \hat{\mathbf{P}}_t^f \mathbf{H}_t' + \mathbf{R}_t)^{-1},
    \end{equation}
    which gives us the analysis update
    \begin{equation} \label{eq:vanilla analysis updates}
        \mathbf{u}_{t,i}^a = \mathbf{u}_{t,i}^f + \hat{\mathbf{K}}_t \big(\mathbf{y}_t + \boldsymbol \epsilon_{t,i} - \mathbf{H}_t\mathbf{u}^f_{t,i} \big), \quad i=1,\ldots,N, 
    \end{equation}
    Under the assumption of the Kalman filter, the forecast ensemble is updated to the analysis ensemble exactly. However, when $\mathcal{M}_t(\cdot)$ is nonlinear, Gaussianity cannot be maintained even ignoring the Monte Carlo error of the estimated covariance matrix \cite{WIKLE20071}. Instead, the EnKF analysis updates correspond to Bayesian "best linear-in-the observations" theory \cite{west2006bayesian}.

    \section{Localization methods}

    \subsection{Motivation for localization}

    As noted, the forecast covariance matrix $\mathbf{P}^f_t$ is typically unknown and must be approximated by the ensemble covariance matrix $\hat{\mathbf{P}}^f_t$. However, $\hat{\mathbf{P}}_t^f$ is rank-deficient whenever $N < p$ and may have high sampling error when the ensemble size $N$ is small, which can cause noisy or unphysical state updates. This is a common setting for the EnKF, as limitations on computational resources often prohibit ensemble sizes greater than $100$. In particular, "spurious correlations" are a concern, where nontrivial correlations are estimated where theory suggests none should exist. Spurious correlations are especially apparent when the state space of interest is a large spatial field.

    To ameliorate the effects of spurious correlations, a variety of "localization" techniques have been proposed in the literature. These techniques seek to filter the noisy correlation estimates from the matrix $\hat{\mathbf{P}}_t$. As the name "localization" implies, many of these techniques utilize a priori knowledge of the correlation structure of a latent spatial field; namely, that spatially-distant locations should be uncorrelated. However, to accommodate complex correlation structures or state vectors without natural distance metrics, some adaptive "localization" methods have been introduced.

    \subsection{Distance-based methods}
    
    Distance-based localization techniques enjoy wide use in the data assimilation literature, owing to their interpretability, ease of implementation, low computational cost, and ability to incorporate prior domain knowledge. However, they are not without downsides. Distance-based methods generally rely on expert input to determine the choice of length-scales, which may be difficult to specify a-priori, or they must be tuned by potentially expensive procedures. Furthermore, they generally ignore any spatial or temporal non-stationarity in correlation structure, assume a monotonic relationship between distance and correlation strength, and exclude the possibility of genuine long-range dynamics. Still, when the implicit assumptions behind their use are not severely violated, their advantages remain especially attractive.

    Many distance-based methods fall into two categories: $\mathbf{B}$ localization and $\mathbf{R}$ localization  \cite{BalanceandEnsembleKalmanFilterLocalizationTechniques}.

    \subsubsection{B localization}
    Methods for $\mathbf{B}$ localization act upon the forecast covariance matrix, which is known as the background error matrix in the weather forecasting literature. They typically aim to reduce spurious correlations by modifying the Kalman gain so that
    \begin{equation} \label{eq:localized kalman gain}
        \hat{\mathbf{K}}_t = \big[\boldsymbol \Gamma \circ (\hat{\mathbf{P}}^f_t \mathbf{H}_t')\big] \big[\boldsymbol \Gamma \circ (\mathbf{H}_t' \hat{\mathbf{P}}_t^f \mathbf{H}_t) + \mathbf{R}_t\big]^{-1}.
    \end{equation}
    Here $\circ$ denotes the Schur (entry-wise) product and $\boldsymbol \Gamma$ is a pre-specified correlation matrix that exhibits decay in spatial correlation. A common choice for $\boldsymbol \Gamma$ is the matrix obtained from the compactly-supported Gaspari-Cohn correlation function with half-width parameter $c$ \cite{GaspariCohn, Houtekamer2001},
     \begin{equation}
        C(r) = \begin{cases} 
            -\frac{1}{4}(\frac{|r|}{c})^5 + \frac{1}{2}(\frac{r}{c})^4 + \frac{5}{8}(\frac{|r|}{c})^3 - \frac{5}{3}(\frac{r}{c})^2 + 1, & 0 \leq |r| \leq c \\
            \frac{1}{12}(\frac{|r|}{c})^5 - \frac{1}{2}(\frac{r}{c})^4 + \frac{5}{8}(\frac{|r|}{c})^3 + \frac{5}{3}(\frac{r}{c})^2 - 5 (\frac{|r|}{c}) + 4 - \frac{2}{3} \frac{c}{|r|}, & c < |r| \leq 2c \\
            0, & 2c < |r| 
         \end{cases} 
     \end{equation}
    Here the input variable $r$ is generally taken to be Euclidean distance. One benefit of $\mathbf{B}$ localization is that the Schur product of a positive-definite matrix and a positive-semi-definite matrix is positive-definite whenever the latter has non-zero rows.

    \subsubsection{$\mathbf{R}$ localization}
    Methods for $\mathbf{R}$ localization instead act upon the observation variance $\mathbf{R}$. They generally inflate the observation variance by a function of physical distance from the individual state components, such that the impact of far-away observations on the analysis update equations is limited.

    \subsection{Adaptive methods}

    To relax the rigid assumptions implicit in distance-based methods of localization, a variety of adaptive methods have been introduced. These methods generally place emphasis on the magnitude of the estimated forecast correlations, although they may also include some spatial information. While the adaptive nature of these methods is alluring, their statistical efficiency is not well-understood, and their performance may hinge on the choice of hyperparameters that are difficult to interpret. Two popular adaptive methods are the hierarchical filter of Anderson (2006) and the ECO-RAP method of Bishop and Hodyss (2009).

    \subsubsection{Hierarchical filter}

    Anderson's hierarchical filter begins by splitting the EnKF ensemble of size $N$ into $L$ sub-ensembles of size $\frac{N}{L}$. At each analysis step, the Kalman gain for each sub-ensemble is computed separately to obtain $\hat{\mathbf{K}}_{t,1}, \ldots, \hat{\mathbf{K}}_{t,L}$. The Kalman gain of each sub-ensemble is shrunken by a measure of disagreement between them. Mathematically, let 
    \begin{equation}
        \kappa_{ij\ell} = \big[\hat{\mathbf{K}}_{t,\ell}\big]_{ij}
    \end{equation}
    for all $i,j = 1, \ldots, p$ and $\ell = 1, \ldots, L$.
    The hierarchical filter collects disagreement measures
    \begin{equation}
        \alpha_{ij} := \underset{\alpha}{\mathrm{argmin}}  \sum_{\ell = 1}^L \sum_{m = 1,m \neq \ell}^L \big(\kappa_{ij\ell} - \alpha\kappa_{ijm}\big)^2
    \end{equation}
    for all $i,j = 1, \ldots, p$. Using the matrix representation $\big[\hat{\boldsymbol \alpha}\big]_{ij} = \hat{\alpha}_{ij}$, the analysis update equations \ref{eq:vanilla analysis updates} are then replaced by:
    \begin{equation}
        \mathbf{u}_{t,i}^a = \mathbf{u}_{t,i}^f + \big[ \hat{\boldsymbol \alpha} \circ \hat{\mathbf{K}}_t \big] \big[ \mathbf{y}_t + \boldsymbol \epsilon_{t,i} - \mathbf{H}_t \mathbf{u}^f_{t,i} \big], \quad i=1,\ldots,N,
    \end{equation}
    
    The idea of Anderson's hierarchical filter is that estimates with high variability may not be stable and should not be trusted.

    \subsubsection{ECO-RAP}

    Ensemble COrrelations Raised to A Power (ECO-RAP) is another adaptive localization method. The idea behind ECO-RAP is to amplify large correlations and penalize small correlations by exponentiation, then localize the Kalman gain according to the weighted coefficients of the Discrete Cosine Transform applied to the exponentiated correlations. ECO-RAP constructs the correlation matrix
    \begin{equation}
        \boldsymbol \Gamma = (\widetilde{\mathbf{S}\mathbf{C}_t^{\odot q}})'(\widetilde{\mathbf{S}\mathbf{C}_t^{\odot q}}),
    \end{equation}
    where the tilde denotes column-wise normalization, the symbol $\odot q$ denotes entry-wise exponentiation by the power $q$, $\mathbf{S}$ is a smoothing matrix and $\mathbf{C}_t$ is the estimated forecast ensemble correlation matrix for the state vector $\mathbf{u}_t$. We let $\mathbf{S} = \mathbf{D}\mathbf{E}$, where $\mathbf{D},\mathbf{E}$ are $p \times p$ matrices with entries chosen to be
    \begin{equation}
    \big[\mathbf{D}\big]_{ij} = \begin{cases} 
        e^{-(\frac{i}{d})^2}, & i = j \\
        0, & i \neq j 
        \end{cases} 
    \end{equation}    

    \begin{equation}
        \big[\mathbf{E}\big]_{ij} = \begin{cases} 
            \frac{1}{\sqrt{p}}, & j = 1 \\
            \sqrt{\frac{2}{p}}\text{cos}(\frac{\pi(2i - 1)(j-1)}{2p}), & j \neq 1
            \end{cases} 
    \end{equation}
    Here $d$ is a parameter to control the degree of spectral smoothing.

    \section{Numerical experiments for localization}

    \subsection{Toy problem: Gaussian process with known covariance}

\begin{figure}[ht]
        \centering
        \subfloat[\centering Sample correlation function.]{{\includegraphics[width=5cm]{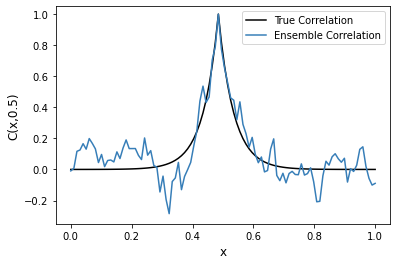} }}%
        \qquad
        \subfloat[\centering Localized correlation function.]{{\includegraphics[width=5cm]{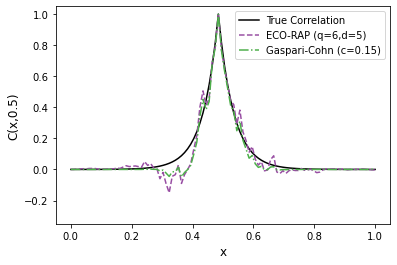} }}%
        \caption{Effects of localization on the estimation of a simple correlation function when the data are Gaussian.}%
        \label{fig:Exponential example}%
    \end{figure}

    We will first illustrate the effects of various localization methods with a simple example where the true correlation function is known analytically. 48 samples were taken of a zero-mean stationary Gaussian process with covariance $C(x,x') = e^{-\frac{|x-x'|}{0.05}}$ over an equally-spaced grid of 101 points on $[0,1]$. Figure \ref{fig:Exponential example}(a) shows the estimated sample correlations over space as an estimate of $C(x,0.5)$. Note that estimation is poor for points that are spatially distant. Figure \ref{fig:Exponential example}(b) shows the effect of applying the Gaspari-Cohn and ECO-RAP localization techniques to the correlation estimates. Estimation at spatially distant points appears to be greatly improved, with many spurious correlations shrinking to zero. The hierarchical filter was not considered for this exposition as that method modifies the entire Kalman gain matrix.

    \subsection{Application to 1-D KPP-Fisher equation: Random diffusivity}

\begin{figure}[ht]
        \includegraphics[width=8cm]{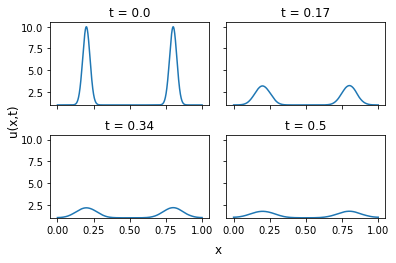}
        \centering
        \caption{The true solution as a function of space at select times.}
        \label{fig:True solution}%
    \end{figure}

\begin{figure}[ht]
        \includegraphics[width=8cm,height=4.8cm]{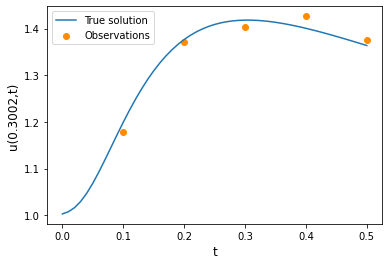}
        \centering
        \caption{The true solution and noisy measurements at the sensor location for a given run of the EnKF.}
        \label{fig:Sensor location}%
    \end{figure}

    To compare localization methods more comprehensively, another study was conducted using the KPP-Fisher equation in one spatial dimension as a forward model, which is a reaction-diffusion equation of the form
    \begin{equation}
        \frac{\partial u}{\partial t} = D \frac{\partial^2 u}{\partial x^2} + u(1 - u).
    \end{equation}
    We modified the equation such that $D = D(x)$ is spatially varying,
    \begin{equation}
        \frac{\partial u}{\partial t} = \frac{\partial{u}}{\partial x} D(x) \frac{\partial u}{\partial x} + u(1 - u).
    \end{equation}
    This partial differential equation was solved using the Finite Volume method. The state space was defined to be the values at 2500 equally-sized cells over the spatial domain $[0,1]$. We followed trajectories over the temporal domain $[0,0.5]$. The initial condition was treated as known and defined to be:
    \begin{equation}
        u(x,0) = 1 + 9e^{-800 (x-0.2)^2} + 9e^{-800 (x-0.8)^2}.
    \end{equation}
    Measurements were taken with with additive $\mathcal{N}(0,\gamma^2)$ noise at a selected cell with center $x = 0.3002$, hereby referred to as the sensor location. We set $\gamma^2 = 0.00025$ to achieve a reasonable ratio of ensemble member variation to measurement noise level. Measurements were taken at times $t = 0.1,0.2,0.3,0.4,0.5$ for a total of five observations. Figure \ref{fig:True solution} shows the true ('ground truth') solution as a function of space for four specific time points, while Figure \ref{fig:Sensor location} illustrates the true solution at the sensor location and an example of measurements taken at that location.

\begin{figure}[ht]
        \centering
        \subfloat[\centering First timestep.]{{\includegraphics[width=5cm]{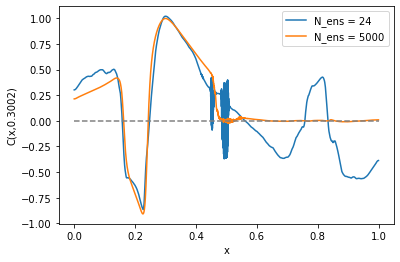} }}%
        \qquad
        \subfloat[\centering Last timestep.]{{\includegraphics[width=5cm]{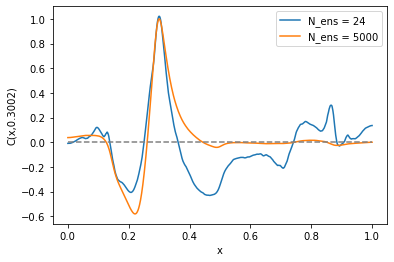} }}%
        \caption{Estimated correlations with the sensor location.}%
        \label{fig:Estimated correlations}%
    \end{figure}

    To initialize our ensemble members, we considered $D$ unknown and assigned a Gaussian process prior to it, $D(\cdot) \sim \mathcal{GP} \big(m,C \big)$. This prior was set with constant mean function $m(x) = 0.005$ and covariance function $C(x,x') = \sigma^2 e^{|x-x'|/0.1}$ where $\sigma = 0.0005$. Furthermore, we set the true $D$ to be a random realization from this prior. Since estimation of $D$ was not of interest, each ensemble member was associated with a realization of $D$ and the ensemble was treated was marginalized over $D$. Figure \ref{fig:Estimated correlations} provides an example of estimated correlations between the cell at the sensor location and other locations in the state space. The poor estimation of spatially-distant correlations suggest the need for localization.

    To assess the performance of localization techniques, we tested the Gaspari-Cohn $\mathbf{B}$ localization method of \cite{Houtekamer2001}, the hierarchical filter of \cite{anderson2007exploring}, and the ECO-RAP method of \cite{bishop2009ecorap1,bishop2009ecorap2}. Initial guesses for the localization tuning parameters were set to be $c = 0.1$ for the Gaspari-Cohn function and $(q,d) = (6,5)$ for ECO-RAP. These parameters were then tuned by considering our model as true, generating a synthetic ground truth for comparison, and doing a trial run of the first step of data assimilation. The criterion chosen was the squared deviation from the ensemble mean and ground truth value (i.e. squared bias), which was averaged over space for a scalar metric. To minimize the effects of sampling variability, we repeated the process 5 times, varying the measurement errors, diffusivities, and measurement perturbations, then averaged the result for a final criterion. A grid search was performed over the set of values $0.05,0.075,0.1,0.125,0.15,0.175,0.2,0.25,$ and $0.3$ for the Gaspari-Cohn half-width parameter and the grid $(q,d) \in \{3,6,9\} \times \{4,5,6\}$ for the ECO-RAP parameters. The parameters $c^* = 0.05$ and $(q^*,d^*) = (9,6)$ were selected for all ensemble sizes based off our criterion, which indicates a strong need for localization. The hierarchical filter was set to $K = 4$ sub-ensembles in accordance with the original paper.
    
\begin{figure}[ht]
        \centering
        \subfloat[\centering First analysis step.]{{\includegraphics[width=5cm]{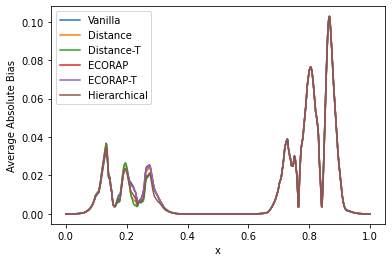} }}%
        \qquad
        \subfloat[\centering Last analysis step.]{{\includegraphics[width=5cm]{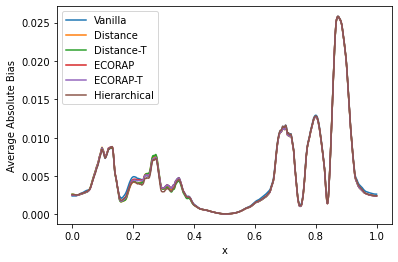} }}%
        \caption{Absolute bias of the EnKF mean averaged over 50 algorithm runs. The ensemble size is fixed at $N = 24$.}%
        \label{fig:Spatial bias plots}%
    \end{figure}

    To mitigate the effect of sampling error, the experiment was repeated 50 times, varying the observation noise, ensemble diffusivites, and EnKF measurement perturbations, while keeping the ground truth the same and all tuning parameters fixed. Only 50 repetitions were conducted due to resource constraints. Figure \ref{fig:Spatial bias plots} shows the absolute bias of the EnKF mean taken as a function of space for $t=0.1$ and $t = 0.5$, averaged over the different runs of the algorithm. We can see that the bias of the EnKF mean is mostly dominated by the dynamics of system, but there is noticeable variation in performance due to localization in the spatial region $[0.15,0.3]$.

\begin{figure}[ht]
        \centering
        \subfloat[\centering After first analysis update.]{{\includegraphics[width=5cm]{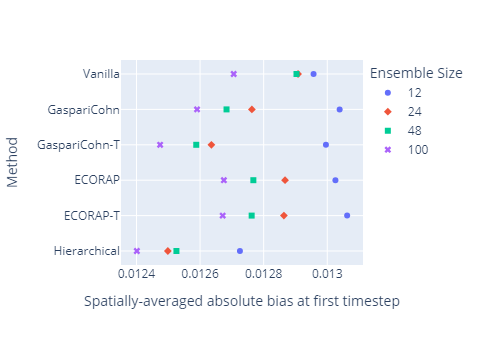} }}%
        \qquad
        \subfloat[\centering After last analysis update.]{{\includegraphics[width=5cm]{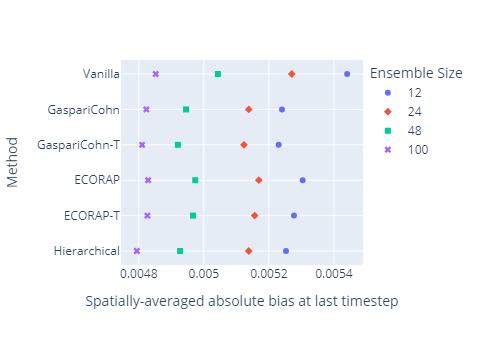} }}%
        \caption{Absolute bias of the EnKF mean averaged over space and 50 algorithm runs.}%
        \label{fig:bias dotplots}%
    \end{figure}

\begin{figure}[ht]
        \centering
        \subfloat[\centering After first analysis update.]{{\includegraphics[width=5cm]{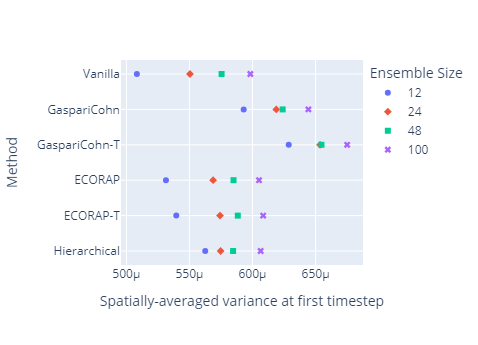} }}%
        \qquad
        \subfloat[\centering After last analysis update.]{{\includegraphics[width=5cm]{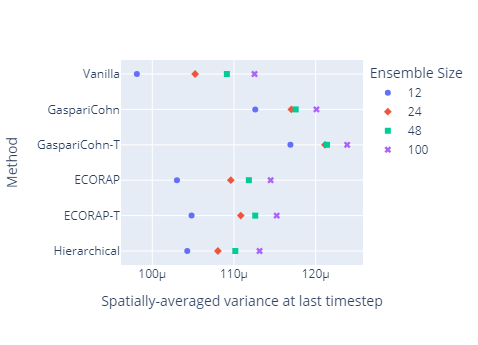} }}%
        \caption{Estimated ensemble variance averaged over space and 50 algorithm runs.}%
        \label{fig:var dotplots}%
    \end{figure}

    Figure \ref{fig:bias dotplots} shows the absolute bias averaged over space for a scalar summary of the EnKF performance. Intuition suggests the necessity of localization methods should wane as the ensemble size increases (and thus the estimation of correlations becomes more precise). We see evidence of this in Figure \ref{fig:bias dotplots}, as the gap between the vanilla and localized EnKF appears to decrease for greater ensemble sizes. In Figure \ref{fig:var dotplots} we see that estimated variances are uniformly higher when using localization methods, which indicates localization can combat filter divergence. Overall, we can see that all localization methods considered offer improvements of the vanilla EnKF in the context of this problem. The hierarchical filter appears to be the preferred choice of localization technique for this problem, with the runner-up being $\mathbf{B}$ localization with Gaspari-Cohn half-width parameter $c^* = 0.05$.

    \section{Computational efficiency of the EnKF}
    Multiple methods exist for the computation of the analysis update in (\ref{eq:analysis updates}). If we collect the ensemble members and model perturbations into matrices 
    \begin{equation}
        \mathbf{u}^a_t = [\mathbf{u}^a_{t,1},\ldots,\mathbf{u}^a_{t,N}],
    \end{equation}
    \begin{equation}
        \mathbf{u}^f_t = [\mathbf{u}^f_{t,1},\ldots,\mathbf{u}^f_{t,N}],
    \end{equation}
    \begin{equation}
        \mathbf{Y}_t = [\mathbf{y}_t + \boldsymbol \epsilon_{t,1},\ldots,\mathbf{y}_t + \boldsymbol \epsilon_{t,N}],
    \end{equation}
    then we can write (\ref{eq:analysis updates}) as
    \begin{equation}
        \mathbf{u}^a_t  = \mathbf{u}^f_t + \hat{\mathbf{P}}^f_t \mathbf{H}_t' \mathbf{Z},
    \end{equation}
    where $\mathbf{Z}$ is the solution to the linear system
    \begin{equation}
        (\mathbf{H}_t \hat{\mathbf{P}}_t^f \mathbf{H}_t' + \mathbf{R}_t) \mathbf{Z} = \mathbf{Y_t} - \mathbf{H}_t \mathbf{u}^f_t,
    \end{equation}
    thus circumventing the inversion of a $q_t \times q_t$ matrix \cite{ninoruiz2015efficientimplementationensemblekalman}. We will call this the 'direct' method (DIR). An alternative method exploits the fact $\mathbf{R}_t$ is often easy to invert or decompose and reduces the problem to inverting an $N \times N$ matrix using the Sherman-Morrison-Woodbury formula (SMW) \cite{mandel2006efficient}. Another approach that only requires the calculation of $\sqrt{\mathbf{R}_t^{-1}}$ is to utilize singular value decompositions (SVD) \cite{mandel2006efficient,ninoruiz2015efficientimplementationensemblekalman}. Lastly, an iterative method utilizing the Sherman-Morrison formula was proposed in \cite{ninoruiz2015efficientimplementationensemblekalman} which only requires the inversion of $\mathbf{R}_t$ (ISM). Theoretical computational complexity for these approaches are available and given in Table \ref{table: theoretical costs}.
    \begin{table}[ht]
        \centering
        \begin{tabular}{ll} \toprule
            {Analysis method} & {Theoretical computational cost} \\ \midrule
            DIR  & $\mathcal{O}(pN^2 + q_t N^2 + q_t^2 N + q_t^3)$ \\
            SMW  & $\mathcal{O}(pN^2 + q_t N^2 + N^3)$ \\
            SVD  & $\mathcal{O}(pN^2 + q_t N^2 + N^3)$ \\
            ISM  & $\mathcal{O}(pN^2 + q_t N^2)$ \\ \bottomrule
        \end{tabular}
        \caption{Theoretical cost of different methods for the analysis update.}
        \label{table: theoretical costs}
    \end{table}
    
    If $\mathbf{R}_t$ is block-diagonal, observations can be assimilated sequentially in independent batches, which only requires inverting square matrices with dimensions of the batch size \cite{Houtekamer2001}. Sequential updates can be combined with any of the previously mentioned techniques. However, we do not consider this option as assimilating observations one-by-one was found to be prohibitively slow.

    A major limitation of the direct method is the need to access the $q_t \times q_t$ matrix $(\mathbf{H}_t \hat{\mathbf{P}}_t^f \mathbf{H}_t' + \mathbf{R}_t)$. Naive storage of this matrix in memory quickly becomes infeasible for $q_t \geq 5e4$. As such, we only investigate this method up until that cutoff. 
    
    Table \ref{table: all methods empirical cost 100 runs} shows empirical computation times for the selected methods when when $p,q$ vary over $5e3,1e4,$ and $5e4$ with $p \geq q$. Computations were performed using Python 3.7.6, utilizing the library numpy 1.18.1 wherever possible, with the exception of scipy version 1.4.1 for the solution linear systems and SVD. The observation operator was assumed to be dense and was applied to the state vector before computations. Observations were taken to be independent, i.e. $\mathbf{R}_t$ is diagonal. As we can see from Table \ref{table: all methods empirical cost 100 runs}, the SVD and SMW analysis update techniques are substantially more computationally efficient than DIR and ISM. The poor performance of ISM, despite the attractive theoretical computational complexity, is likely due to the method's sequential Python implementation which does not take advantage of numpy's vectorized operations. Table \ref{table: two methods empirical cost 100 runs} demonstrates the superior performance of the SMW method up until the largest state and observation space considered, where SVD is estimated to perform slightly better but with larger variance in computation times.

    \begin{table}[ht]
        \centering
        \begin{tabular}{@{\extracolsep{4pt}}llrrrrrrrr}
        \toprule   
        {} & {State Dimension} & \multicolumn{3}{c}{Observation Count}\\
        \cmidrule{2-2}
        \cmidrule{3-5} 
         Method & p & 5e3 & 1e4 & 5e4 \\ 
        \midrule
        DIR & 5e3 & 3075.4 $\pm$ 23.0 & & \\
        SMW &  & $\mathbf{56.4 \pm 3.1}$ &  & \\ 
        SVD &  & 109.2 $\pm$ 8.5 &  & \\ 
        ISM &  & 2073.3 $\pm$ 51.0 &  & \\ 
        DIR & 1e4 & 3106.8 $\pm$ 14.4 & 10573.1 $\pm$ 22.8 & \\
        SMW &  & $\mathbf{72.7 \pm 4.6}$ & $\mathbf{116.2 \pm 6.1}$ & \\  
        SVD &  & 121.9 $\pm$ 7.2 & 148.7 $\pm$ 5.4 & \\ 
        ISM &  & 2063 $\pm$ 88.9 & 3996.5 $\pm$ 103.1 & \\ 
        DIR & 5e4 & 3189.8 $\pm$ 20.3 & 10624.1 $\pm$ 25.3 & ---\\
        SMW &  & $\mathbf{159.1 \pm 6.5}$ & $\mathbf{209.5 \pm 9.8}$ & $\mathbf{774.7 \pm 10.1}$\\
        SVD &  & 205.1 $\pm$ 11.3 & 236.6 $\pm$ 11.8 & 850.0 $\pm$ 34.1\\  
        ISM &  & 2168.8 $\pm$ 109.7 & 4192.6 $\pm$ 98.1 & 10786.1 $\pm$ 141.8\\
        \bottomrule
        \end{tabular}
        \caption{Mean $\pm$ standard deviation times for 10 runs averaged over 100 loops for each of analysis update formulations. Units are in miliseconds. Ensemble size was set to $N = 100$ for all calculations.}
        \label{table: all methods empirical cost 100 runs}
\end{table}

    \begin{table}[ht]
        \centering
        \begin{tabular}{@{\extracolsep{4pt}}lllll}
        \toprule   
        {} & {State Dimension} & \multicolumn{3}{c}{Observation Count}\\
        \cmidrule{2-2}
        \cmidrule{3-5} 
         Method & p & 5e5 & 1e6 & 5e6 \\ 
        \midrule
        SMW & 5e5 & $\mathbf{1.16 \pm 0.01}$ & & \\
        SVD &  & 2.20 $\pm$ 0.02 &  & \\ 
        SMW & 1e6 & $\mathbf{1.29 \pm 0.20}$ & $\mathbf{2.13 \pm 0.02}$ & \\
        SVD &  & 2.33 $\pm$ 0.24 & 4.23 $\pm$ 0.12 & \\ 
        SMW & 5e6 & $\mathbf{2.36 \pm 0.01}$ & $\mathbf{3.20 \pm 0.01}$ & 41.7 $\pm$ 0.37 \\
        SVD &  & 3.26 $\pm$ 0.64 & 5.19 $\pm$ 44.4 & $\mathbf{40.2 \pm 1.33}$\\
        \bottomrule
        \end{tabular}
        \caption{Mean $\pm$ standard deviation times for 10 runs averaged over 100 loops for each of the two cheaper analysis update formulations. Units are in seconds. Ensemble size was set to $N = 50$ for all calculations.}
        \label{table: two methods empirical cost 100 runs}
\end{table}

\section{Conclusion}
Ensemble Kalman methods are extremely powerful and cost-effective tools for combining physics-based models with real-world measurements to provide estimates of a state variable. They can provide good performance on large-scale systems even with less than one-hundred forward model runs. However, as we have shown, even in non-Numerical Weather Prediction contexts the ensemble Kalman filter is still subject to spurious correlations and can benefit  from localization methods. Specifically, we recommend the Gaspari-Cohn localization when a natural distance metric exists and Anderson's hierarchical filter when no such metric exists or is appropriate to the dynamics of the problem. Furthermore, we note that the analysis update can be expedited considerably using the Sherman-Morrison-Woodbury formula.

\section*{Acknowledgements}

Sandia National Laboratories is a multimission laboratory managed and operated by National Technology and
Engineering Solutions of Sandia , LLC., a wholly owned subsidiary of Honeywell International, Inc., for the U.S.
Department of Energy’s National Nuclear Security Administration under contract DE-NA-0003525. This paper
describes objective technical results and analysis. Any subjective views or opinions that might be expressed in the
paper do not necessarily represent the views of the U.S. Department of Energy or the United States Government.
This article has been authored by an employee of National Technology \& Engineering Solutions of Sandia,
LLC under Contract No. DE-NA0003525 with the U.S. Department of Energy (DOE). The employee owns all
right, title and interest in and to the article and is solely responsible for its contents. The United States Gov-
ernment retains and the publisher, by accepting the article for publication, acknowledges that the United States
Government retains a non-exclusive, paid-up, irrevocable, world-wide license to publish or reproduce the published
form of this article or allow others to do so, for United States Government purposes. The DOE will provide
public access to these results of federally sponsored research in accordance with the DOE Public Access Plan
https://www.energy.gov/downloads/doe-public-access-plan.

\bibliographystyle{siam}
\bibliography{proceedings}

\end{document}